\documentstyle[11pt,fleqn]{article}
\def\titolo{\par\bigskip\begin{center}\bf\LARGE}
\def\endtitolo{\end{center}\par\bigskip\par\rm\normalsize}
\def\instit{\begin{center}\large}
\def\endinstit{\end{center}\rm\normalsize}
\def\references{\begin{thebibliography}{10}}
\def\endreferences{\end{thebibliography}\end{document}}

\newcommand{\btit}{\begin{titolo}}
\newcommand{\etit}{\end{titolo}}

\renewcommand{\author}[1]{\begin{center}\Large #1\end{center}}
\renewcommand{\date}[1]{\par\bigskip\par\sl\hfill #1\par\medskip\par}

\newcommand{\pacs}[1]{\smallskip\noindent{\sl PACS number(s):
                       \hspace{0.3cm}#1}\par\bigskip}
\newcommand{\babs}{\hrule\par\begin{description}\item{Abstract: }\it}
\newcommand{\eabs}{\par\end{description}\hrule\par\medskip\rm}
\newcommand{\ack}[1]{\par\section*{Acknowledgments} #1}
\renewcommand{\vec}[1]{{\bf #1}}       
\newcommand{\ca}[1]{{\cal #1}}         
\newcommand{\nn}{\nonumber}            
\newcommand{\beq}{\begin{eqnarray}}    
\newcommand{\eeq}{\end{eqnarray}}      
\newcommand{\beqn}{\begin{eqnarray}}   
\newcommand{\eeqn}{\end{eqnarray}}     
\newcommand{\at}{\left(}               
\newcommand{\aq}{\left[}               
\newcommand{\ag}{\left\{}              
\newcommand{\ct}{\right)}              
\newcommand{\cq}{\right]}              
\newcommand{\cg}{\right\}}             
\newcommand{\N}{\mbox{$I\!\!N$}}                 
\newcommand{\Z}{\mbox{$Z\!\!\!Z$}}               
\newcommand{\C}{\mbox{$I\!\!\!\!C$}}             
\newcommand{\ii}{\infty}                         
\newcommand{\fr}[2]{\mbox{$\frac{#1}{#2}$}}      
\newcommand{\Tr}{\,\mbox{Tr}\,}                  
\renewcommand{\Im}{\,\mbox{Im}\,}                

\newcommand{\be}{\beta}

\newcommand{\ep}{\varepsilon}
\newcommand{\ze}{\zeta}

\newcommand{\la}{\lambda}

\newcommand{\th}{\theta}

\newcommand{\Ga}{\Gamma}

\begin{document}
{\LARGE \bf
On the high-temperature behaviour of the closed superstring }\\

\vspace{5mm}

{\sc S. Dusini} \\ {\it Department of Physics, University of Trento, 38050
Povo,
Italy} \\
{\sc S. Zerbini} \\
{\it Department of Physics, University of Trento, 38050 Povo,
Italy}    \\
{\it I.N.F.N., Gruppo Collegato di Trento}

\vspace{5mm}

\babs
The high-temperature expansion  for closed super-string one-loop free energy
is studied. The Laurent series representation is obtained and its sum
is analytically continued in order to investigate the nature of the  critical
(Hagedorn) temperature. It is found that beyond this critical
temperature the statistical sum contribution of the free energy
is finite but has an imaginary part, signalling a possible metastability of the
system.
\eabs

\vspace{8mm}


 The interest in  (super)-strings at
non-zero temperature  has been recently grown (see for example Refs.
[1-6]).
One of the main reasons for these investigations is related to the
 thermodynamics  of the early universe (see [7] and references therein) as
well as to the attempts  to make use of strings  for the
description of the high-temperature limit of
the confining phase of large-N SU(N) Yang-Mills theory [8,9].

Since the early days of dual string models it is  known that such extended
objects have an exponentially dependence
on the mass of the asymptotic state level density.  This leads to the
existence of the Hagedorn temperature [1] and the breakdown  above
this critical temperature
of the correspondence between canonical microcanonical ensemble.
We also mention  that the Hagedorn temperature
is thought to be the critical temperature for a first order phase transition
with a large latent heat [3].

In a series of recent papers  [11,12],  making use of the so-called
Mellin-Barnes
representation for the one-loop open superstring free energy, the high
temperature behaviour for open and closed bosonic string and open
superstring has been investigated. In particular, the Hagedorn
temperatures have been characterized as radii of convergence of the
associated Laurent series representations.
The aim of this note is to
complete the analysis to the closed superstring, which, besides to be
free from infrared divergences, which plague the bosonic strings,
presents some peculiarity. The most important one being the existence of
the free energy at the critical temperature, in contrast to the
open cases, where the free energy has a pole singularity.
As a consequence, the derivation of the related Laurent representation
and its analytical continuation seems worth being investigated.

To begin with, we recall that there exist several representations  for the
one-loop string free energy. One of these
representations [6] gives a modular-invariant expression for the free
energy. However, this and all the other well known representations
 [5] are {\it integral} ones in which the Hagedorn
temperature appears as the  convergence condition in the
ultraviolet limit of a certain integral.
Here we shall make use of the well-known approach to finite
temperature field theory and then we shall generalize it to
the superstring case.

We recall that if we are dealing with bosons (fermions) at finite
temperature ($\be$ being the inverse of the temperature) in a D-dimensional
space-time, we may
consider the fields  on $S^1\times M^{D-1}$, the imaginary time variable
$\tau=it$ being compactified with  boson (fermion) fields assumed to
be periodic (anti-periodic) in $\be$. The related one-loop partition function
may be written (we consider scalar field $\phi$, the spinor case can be treated
similarly)
\beq
Z_\be= \int [d \phi] e^{-\fr{1}{2}\int_0^\be d\tau \int dx \phi L_\be
\phi}\,,
\eeq
where $L_\be=-\partial_\tau^2+\vec {p}^2+m^2$.
The related free energy reads
\beq
\beta \ca{F(\be)}=-\log Z_\be=\frac{1}{2}\log \det L_\be=
-\frac{1}{2}\int_{0}^{\ii}dt t^{\ep-1}\Tr e^{-t L_\be}\,.
\label{fe}
\eeq
In the above equation, the determinant of the differential operator
$L_\be$ has been regularized by introducing the ultraviolet cut-off
$\ep$.
For scalar fields we have
\beq
\Tr_b  e^{-t L_\be}=\frac{V_{D-1}}{(4 \pi t)^{\fr{D-1}{2}}}
\sum_{\Z}e^{-\fr{4\pi^2 n^2 t}{\be^2}}e^{-tm^2}\,.
\eeq
 The corresponding fermionic contribution reads
\beq
\Tr_f  e^{-t L_\be}=\frac{V_{D-1}}{(4 \pi t)^{\fr{D-1}{2}}}
\sum_{\Z}e^{-\fr{\pi^2 (2n+1)^2 t}{\be^2}}e^{-tm^2}\,.
\eeq

Adding the two terms, using the Poisson resummation
formula and separating the vacuum term ($n=0$), which
does not depend on $\be$,
one arrives at the following (supersymmetric) statistical sum contribution to
the free energy density ($D=10$)
\beq
F(\be)=F_b(\be)+F_f(\be)=-\frac{1}{2}\int_{0}^{\ii}dt t^{\ep-1}(4\pi
t)^{-5}e^{-t m^2}\aq\th_3\at 0|\frac{i\be^2}{4\pi t}\ct-
\th_4\at 0|\frac{i\be^2}{4\pi t}\ct\cq,
\label{elft}
\eeq
where  $\th_3(x|y)$ and $\th_4(x|y)$ are
two of the Jacobi elliptic theta-functions and $\ep$ acts as an analytic
regularization parameter and we  will take the
limit $\ep  \to 0$ at the end of the calculations.

In order to obtain the free energy in the case of the super-string, one may
simply observe that
the mass $m$ becomes an operator $M$. As a
consequence, the expression for the statistical sum
(\ref{elft}) takes the form
\beq
F(\be)=-\frac{1}{2}\int_{0}^{\ii}dt t^{\ep-1}(4\pi
t)^{-5}\Tr e^{-t M^2}\aq\th_3\at 0|\frac{i\be^2}{4\pi t}\ct-
\th_4\at 0|\frac{i\be^2}{4\pi t}\ct\cq
\label{elstringa}
\eeq
where $M^2$ is the mass operator.  For the closed superstring in
the light cone gauge ($T=\frac{1}{\pi}$) one has (see [13])
\beq
M^2=4\sum_{i=1}^{8}\sum_{1}^{\ii}n\at
N_{n_i}^b+N_{n_i}^f+\tilde N_{n_i}^b+\tilde N_{n_i}^f\ct.
\label{massassc}
\eeq
Furthermore in this case (closed superstring) we also have to take into
account the following
constraint on the states of the system [13]
\beq
\sum_{i=1}^{8}\sum_{n=1}^{\ii}n\at N_{n_i}^b+N_{n_i}^f-\tilde
N_{n_i}^b-\tilde N_{n_i}^f\ct=0
\label{N-N}
\eeq
This constraint, which reflects the absence of a preferred point on
the closed string, may be implemented via the usual identity (see for
example [5]). Therefore the trace of the
"heat-kernel" of the mass operator becomes
\beq
\Tr\ag e^{-M^2t}\int_{-\frac{1}{2}}^{\frac{1}{2}}d\tau_1 e^{2\pi
i\tau_1(N-\tilde
N)}\cg &=& 64\int_{-\frac{1}{2}}^{\frac{1}{2}}d\tau_1\prod_{n=1}^{\ii}\left |
\frac{1-e^{2\pi i\tau n}}{1+e^{2\pi i\tau n}}\right
|^{-16}\nn\\
&=& 64\int_{-\frac{1}{2}}^{\frac{1}{2}}d\tau_1|\th_4(0|2\tau)|^{-16}
\label{traccia}
\eeq
where $\tau=\tau_1+i\tau_2=\tau_1+i\frac{2t}{\pi}$ and the
presence of the factor 64 is due to the degeneracy of the ground
states.

Using the
asymptotic behavior of the $\th_4$   for $\tau_1=0$ and
$\tau_2\rightarrow 0$, the integrand of Eq. (\ref{elstringa})
behaves, in this limit, as
\beq
{\th_4(0|2\tau)^{-16}}_{\Big |_{\tau_1=0,\tau_2\rightarrow 0}}=
\frac{\tau_2^8}{2^8}\at e^{\frac{2\pi}{\tau_2}}-16e^{\frac{\pi}{\tau_2}}+120
+O\at e^{-\frac{\pi}{\tau_2}}\ct\ct
\label{th4asin}
\eeq
Thus, in order to isolate the high-temperature behaviour we may rewrite
identically
\beq
\Tr e^{-M^2t}=64\int_{-\frac{1}{2}}^{\frac{1}{2}}d\tau_1
\aq |\th_4(0|2\tau)|^{-16}-\frac{\tau_2^8}{2^8}\at e^{\frac{2\pi}{\tau_2}}-
16e^{\frac{\pi}{\tau_2}}+120\ct\cq\nn\\
+\frac{\tau_2^8}{2^8}\at
e^{\frac{2\pi}{\tau_2}}-16e^{\frac{\pi}{\tau_2}}+120\ct.
\label{tr reg}
\eeq
{}From the expression (\ref{tr reg}) and with the help of the Poisson
summation formula,  the asymptotic behavior  at high temperature of the
statistical sum (\ref{elstringa}) becomes
\beqn
F(\be) &\simeq & -\frac{1}{8\pi^7}\at\frac{\ln 2}{3!}+
G(1)\frac{2\pi}{\be}\ct\nn\\
&-& \frac{4}{2^8 \pi^{10}} \int_{0}^{\ii}d\tau_2\tau_2^{\ep +2}
\at e^{\frac{2\pi}{\tau_2}}-16e^{\frac{\pi}{\tau_2}}+120\ct
\sum_{n=0}^{\ii}\exp\at -\frac{(2n+1)^2\be^2}{2\pi\tau_2}\ct
\label{ab}
\eeqn
where we have neglected those terms that are exponentially small at
high temperature ( $\be \to 0$) and have put
\beq
G(s)=\frac{32}{\pi^3\sqrt{2\pi}}\int_{0}^{\ii}d\tau_2\tau_2^{\frac{s}{2}-6}
\aq\int_{-\frac{1}{2}}^{\frac{1}{2}}d\tau_1 |\th_4(0|2\tau)|^{-16}-
\frac{\tau_2^8}{2^8}\at
e^{\frac{2\pi}{\tau_2}}-16e^{\frac{\pi}{\tau_2}}+120\ct\cq.
\label{G(s)}
\eeq
As a result $G(1)$ is a finite number. The second term of (\ref{ab})
is the one which interests us, since it describes the Hagedorn sector.
In fact it is easy to see that the integral converges as soon as $
\be^2 > 4\pi^2$, $2\pi$ being the Hagedorn critical temperature:
this is a well known result (see for example Ref. [14]). However, our
aim is to find an analytical continuation of this expression.
 To evaluate it, we expand
$e^{\frac{2\pi}{\tau_2}}$ and $e^{\frac{\pi}{\tau_2}}$ in series of powers and
then we integrate terms by terms. The integrals thus obtained,
have to be
analytically regularized, and they can be evaluated with the help of
the relation
\beq
\int_{0}^{\ii}dy
y^{k-4+\ep}\exp\at-\frac{(2n+1)^2\be^2y}{2\pi}\ct=\Ga(k-3+\ep)
\at\frac{(2n+1)^2\be^2}{2\pi}\ct^{3-k-\ep}
\eeq
which has to be interpreted in the sense of analytic continuation. Moreover the
regularization of the integral $\int_{0}^{\ii}t^{\la}dt$ as analytical
function of $\la$ gives $\int_{0}^{\ii}t^{\la}dt=0$ [15], and therefore the
last integral in the second term of (\ref{ab}) gives no contribution
to the statistical sum.
The sum over $n$ gives the series representation of the $\ze$-function.
The removal of the cutoff is delicate. Again the analytical
continuation must be invoked. The first four terms in the sum are
treated
by making use of functional equation for the Riemann $\ze$-function,
namely
\beq
\Ga(z)\ze(2z)=  \Ga(\fr{1}{2}-z)\ze(1-2z)\pi^{2z-1}\,.
\eeq
As a result, by taking the limit for $\ep$ that goes to zero
we end up with a representation for the asymptotic
behavior at high temperature of the free energy density in terms of  the
Laurent series
\beq
F(\be)\simeq -\frac{1}{8\pi^7}\aq\frac{\ln 2}{3!}+G(1)x+\sum_{k=0}^{2}
A(k)x^{2k-6}+\sum_{k=1}^{\ii}B(k)x^{2k}\cq
\label{el laurent}
\eeq
where $x=\frac{2\pi}{\be}$ and
\beq
A(k) &=& \frac{\pi^{2k-7}}{k!}(1-2^{6-2k})\Ga\at\frac{7}{2}-k\ct
\ze\at 7-2k \ct\nn\\
B(k) &=& (1-2^{1-k})(1-2^{-2k})\ze(2k)\frac{\Ga(k)}{(k+3)!}.
\label{AABB}
\eeq
One may derive this Laurent representation starting from the
Mellin-Barnes representation along the lines of Refs. [11,12]. Here
we have presented another derivation.

As we have already mentioned, this series
representation characterizes the critical temperature. In fact the series
converges when $\be\geq\be_c=2\pi$, $\be_c$ being the
Hagedorn temperature.
It is interesting to notice that, unlike the case of the open superstring,
the free energy of the closed superstring is finite for
$\be\rightarrow\be_c+0$ (see for example the review papers [5] and
[14]).

Inside the
 radius of convergence, it is possible to evaluate the sum of the series
$\sum_{k=1}^{\ii}B(k)x^{2k}$ in terms of known functions. The
calculation is tedious but straightforward and we only sketch the
derivation.
The starting point is the relation (here $x <1$)
\beq
\sum_{k=1}^\ii (1-2^{-2k})\ze(2k)\frac{x^{2k+1}}{2k}=-\frac{x}{2}\ln
\cos (\frac{\pi x}{2})\,.
\eeq
Integrating terms by terms one gets
\beq
\sum_{k=1}^\ii (1-2^{-2k})\ze(2k)\frac{x^{2k+2}}{2k(2k+2)}=-\frac{1}{2}
\int_0^x dt t\ln \cos (\frac{\pi t}{2})\,.
\eeq
Now one has [16]
\beq
\int dt \ln \cos t=\frac{i}{2}Li_2(-e^{2it})-t \ln 2-\frac{i}{2}t^2
\eeq
where  $Li_s(z)$ is the polylogarithm function of order $s$. It
is an analytic function for $s,z\in\C$ and $|z|<1$, defined by the
Dirichelet series (see Ref. [17], where this function has been
investigated in detail)
\beq
Li_s(z)=\sum_{n=1}^{\ii}\frac{z^n}{n^s}\qquad |z|<1\ s\in\C.
\label{def Li}
\eeq
For $s\in\N\ s\geq 2$ the function is also defined for $z=1$ and in
this case it is equal to the Riemann $\ze$-function.
Making use of the following property of the polylogarithm function
\beq
\int_0^y Li_s(-e^{iat})dt=\frac{1}{ia}\aq
Li_{s+1}(-e^{iay})-Li_{s+1}(-1)\cq
\eeq
and
integrating twice terms by terms, one arrives at the final result, which reads
\beq
\sum_{k=1}^{\ii}B(k)x^{2k}=\frac{4i}{\pi^2 x^3}\tilde{L}i_4\at -e^{i\pi x}\ct
-\frac{24}{\pi^3 x^4}\tilde{L}i_5\at
-e^{i\pi x}\ct -\frac{60i}{\pi^4 x^5}\tilde{L}i_6\at -e^{i\pi x}\ct\nn\\
+\frac{60}{\pi^5 x^6}\tilde{L}i_7\at -e^{i\pi x}\ct
-\frac{60}{\pi^5 x^6}\tilde{L}i_7(-1)
-\frac{6}{\pi^3 x^4}\tilde{L}i_5(-1)-\frac{1}{2\pi x^2}\tilde{L}i_3(-1)\nn\\
+\frac{4\pi}{48}\ln2 +\frac{16i\pi^2}{840}x(1-2^{-\frac{1}{2}})
\label{somma}
\eeq
where we have introduced the functions
\beq
\tilde{L}i_n(-e^{i\pi x})=Li_n(-e^{i\pi
x})-2^{\frac{n+1}{2}}Li_n(-e^{i\pi\frac{x}{\sqrt{2}}})
\label{du}
\eeq

Let us conclude with some remarks. Eq. (\ref{somma})  has been obtained for
$x \leq 1$. However, the function
$Li_s(z)$ admits an analytical continuation in the cut z-plane, the cut
being imposed from $1$ to $+\ii$ along the positive real axis [17]. A
direct calculation shows that the
function (\ref{somma}) is real for  $x \leq 1$. For $x >1$, namely for
a temperature bigger than the Hagedorn one, the function is finite
but it acquires a non-vanishing imaginary part. The idea to use the
analytical continuation in string theory is not new. It can be found
for example, in Ref. [18], where the appearance  of imaginary terms
in the free energy above the Hagedorn temperature has been pointed out. Note,
however, that the use of the
analytical continuation may present some problems (see the criticism
contained in Ref. [19]).

Having the explicit form of the statistical sum, the imaginary part
can be computed and the result is
\beq
\Im F_\be \equiv \frac{\pi^2}{196}\aq \theta(x-1)(x^2-1)^3
-\theta(x-\sqrt 2)\frac{(x^2-2)^3}{4}\cq
\eeq
This is not surprising, because naive considerations based on the
exponential  dependance on the mass of the state  level density leads
to the conclusion that above the critical temperature the one-loop
partition function is diverging. Since  the analytical
continuation of the free energy  has a non-vanishing imaginary part, in
analogy with field theoretical case [20] and finite temperature
one-loop quantum gravity [21], one might conclude that the
"effective potential" for superstring, above the Hagedorn temperature,
develops a kind of  new  "local minimum" and there might be a decay from the
old vacuum to the new one [18], the decay rate of this "tunneling" being
described by the imaginary part we have computed.
One might conclude that above the
Hagedorn
temperature, the superstring would become a metastable system. However,
it should be
noted that we are neglecting  gravitational effects and this
could not be justified  when the temperature approches the Hagedorn one [13].
Finally similar considerations may be done for the heterotic string
[22] and we hope to return on this issue elsewhere.

\ack{ We would like to thank A.A. Bytsenko and S.D. Odintsov for useful
discussions}

\end{document}